\documentclass[]{interact}

\usepackage{epstopdf}
\usepackage{subfigure}

\usepackage[numbers,sort&compress,merge]{natbib}
\bibpunct[, ]{(}{)}{,}{n}{,}{,}
\renewcommand\bibfont{\fontsize{10}{12}\selectfont}
\renewcommand\citenumfont[1]{\textit{#1}}
\renewcommand\bibnumfmt[1]{(#1)}

\theoremstyle{plain}

\theoremstyle{definition}

\theoremstyle{remark}

\begin{document}
\articletype{ARTICLE TEMPLATE}
\title{Efficient eigenvalue determination for arbitrary Pauli products based on generalized spin-spin interactions}
\author{
\name{D. Leibfried\textsuperscript{a}\thanks{CONTACT D. Leibfried Email: dil@boulder.nist.gov}  and D.J. Wineland\textsuperscript{a}}
\affil{\textsuperscript{a} Ion Storage Group, Time and Frequency Division, National Institute of Standards and Technology, Boulder, CO 80305, USA}
}
\maketitle
\begin{abstract}
Effective spin-spin interactions between $N+1$ qubits enable the determination of the eigenvalue of an arbitrary Pauli product of dimension $N$ with a constant, small number of multi-qubit gates that is independent of $N$ and encodes the eigenvalue in the measurement basis states of an extra ancilla qubit. Such interactions are available whenever qubits can be coupled to a shared harmonic oscillator, a situation that can be realized in several physical qubit implementations. For example, suitable interactions have already been realized for up to 14 qubits in ion traps. It should be possible to implement stabilizer codes for quantum error correction with a constant number of multi-qubit gates, in contrast to typical constructions using a number of two-qubit gates that increases as a function of $N$. The special case of finding the parity of $N$ qubits only requires a small number of operations that is independent of $N$. This compares favorably to algorithms for computing the parity on conventional machines, which implies a genuine quantum advantage.
\end{abstract}
\vspace{0.5 cm}
\begin{center}
{\it We dedicate this work to Danny Segal, scientist, community builder and friend. Your bold and cheerful way of leaving the trodden paths and pursuing your own goals have made our field richer. We will not be able to fill your void, but we will carry on.}\\
\end{center}
\section{Preface}
Some of the work presented here was developed in 2004 to support a spectroscopy demonstration with three entangled ions \cite{leibfried04}, where a special case of the key relation Eq.(\ref{Eq:UniTra}) below was stated without proof and without discussing further implications. This publication appeared before supplemental materials became common and the proof was too long to include in a letter-style publication at the time. Thirteen years later, we have picked up the thread again and generalize the special case stated in \cite{leibfried04} to relate more general spin-spin interactions to arbitrary Pauli product operators.\\
\\
It may be rare that experimental physicists resort to the principle of mathematical induction, as we have done below. However, our contribution seems to be in tune with a particularly important  lesson that Danny's life can teach us: You should not be afraid to do things in your own way, no matter how far off the beaten path this might take you.
\section{Generalized spin-spin interactions and Pauli products}
We consider $N$ two-level systems with
    logical basis of the $l$-th qubit ($l ~\epsilon~ \{1,...,N\}$) defined by the eigenstates $\{| \downarrow \rangle, | \uparrow \rangle\}$ of the $z$-component of a spin-1/2 angular momentum operator $S_{l,k}$ with $k=\{x,y,z\}$. We write $\sigma_{l,i}$ for the $2\times2$ identity matrix in the state space of the $l$-th qubit  and $\sigma_{l,k}$ for the Pauli-matrices for $k=\{x,y,z\}$. When setting $\hbar=1$ for simplicity, the eigenvalues in the measurement basis are
\begin{equation}
S_{l,z} | \uparrow \rangle={\textstyle \frac{1}{2}} \sigma_{l,z} | \uparrow \rangle = {\textstyle \frac{1}{2}}|
\uparrow \rangle,
S_{l,z}| \downarrow \rangle={\textstyle \frac{1}{2}} \sigma_{l,z} | \downarrow \rangle = -{\textstyle \frac{1}{2}}|
\downarrow \rangle.
\end{equation}
 Our previous work relied on the collective angular momentum operator in the $N$-spin Bloch vector representation 
\begin{equation}\label{Eq:DikRep}
\vec{J} = \sum_{l=1}^N \vec{S}_l=1/2 \sum_{l=1}^N \vec{\sigma}_l.
\end{equation}
This notation allows, for example, for a compact representation of the collective rotation operator $R^{(N)}_k \equiv \exp[-i {\textstyle
\frac{\pi}{2}} J_k]$ ($k=\{x,y,z\}$), often called a ``$\pi/2$"-pulse, applied to all $N$ qubits  uniformly.\\ 
\\
M{\o}lmer and S{\o}rensen  \cite{molmer99,sorensen00} (for $k=\phi$ and $\sigma_{l,\phi}=\cos(\phi) \sigma_{l,x}-\sin(\phi) \sigma_{l,y}$) and Milburn  \cite{milburn00} (for $k=z$) showed how the interaction 
\begin{equation}\label{Eq:SquOpe}
U_k = \exp[-i\chi J_k^2],
\end{equation}
that acts uniformly on all qubits with a suitable coupling parameter $\chi$  can be implemented in trapped ion systems (see also \cite{solano99}). The approach in ion traps can be generalized to any system of qubits that couples uniformly to one harmonic oscillator.  The theoretical work set off a flurry of experiments in which entangled states of two to 14 ion qubits were eventually produced based on such $J^2$-interactions. For example, work described in \cite{leibfried04,sackett00,meyer01,leibfried03,leibfried05,benhelm08,friedenauer08,kim10,monz11} produced entangled states by using lasers to couple the qubits to a harmonic normal-mode motion of the ions and in \cite{ospelkaus11,khromova12,weidt16,harty16} this was accomplished with microwaves. These and other similar experiments encompassed an wide range of contexts, including precision spectroscopy, quantum information processing, and quantum simulation.\\ 
\\
Here we generalize $U_k$, in which interactions are uniformly along direction $k$ over all qubits, to allow for interactions along different directions on different qubits. In particular, for every product of Pauli matrices and the identity over $N$ qubits ($k_l=\{i,x,y,z\}$),
\begin{equation}\label{Eq:StaProd}
P_N = \prod_{l=1}^N \sigma_{l,k_l},
\end{equation}
we define an associated operator $D_N$ 
\begin{equation}\label{Eq:StaSum}
D_N = 1/2 \sum_{l=1}^N \sigma_{l,k_l},
\end{equation}
that can be used as the generator of a generalized spin-spin $D_N^2$-operation according to $G_N=\exp[-i \alpha D_N^2]$. The connection between $P_N$ and $D_N$ will be discussed in section \ref{Sec:ConBet}. Products of Pauli-matrices like $P_N$ are essential as so called stabilizers in a certain class of quantum-error correction codes \cite{nielsen10} (see section \ref{Sec:AppErr}).  In the special cases $k_l=k$ for all $l$, the operator $D_N$ is equivalent to $J_k$, therefore all relations that hold for $D_N$ will also be true for any $J_k$ of dimension $N$.
\section{Connection between generalized spin-spin interactions and Pauli product operators}\label{Sec:ConBet}
With the above definitions, we now state our main result: Depending on whether $N$ is odd or even, the corresponding $D_N$ and $P_N$ fulfill
\begin{eqnarray}\label{Eq:UniTra}
U_N=\left\{
\begin{array}{ll}
\exp[-i \textstyle{\frac{\pi}{2}} D_N^2]&;\ N\ {\rm even}\\
\exp[-i \textstyle{\frac{\pi}{2}} D_N]
    \exp[-i \textstyle{\frac{\pi}{2}} D_N^2]&;\ N\ {\rm
    odd}
\end{array} \right. \nonumber \\
    = \frac{\exp(-i \textstyle{\frac{\pi}{4 E}} )}{\sqrt{2}}\left(1+ i^{N+E}
   P_N  \right)
\end{eqnarray}
with $E=1$ for $N$ even and $E=2$ for $N$ odd. Eq. (\ref{Eq:UniTra}) was stated without proof for the uniform cases where $k_l=k$ as Eq. (2) in \cite{leibfried04}. In the special case of $k_l=k=x$ and applying $U_N$ to the initial state $|\downarrow, N\rangle \equiv |\downarrow, \downarrow,...,\downarrow \rangle$, relation (\ref{Eq:UniTra}) was also given by S\o rensen and M\o lmer \cite{molmer99}. The fact that Eq. (\ref{Eq:UniTra}) applies to arbitrary input states is crucial for the efficient algorithms we present in section \ref{Sec:AppErr}, including a constant depth algorithm for finding the parity of a state. It should also be mentioned that since Ref.  \cite{leibfried04} appeared, another related constant depth algorithm for finding the parity was published \cite{zeng05}. That work did not contain a proof of Eq. ({\ref{Eq:UniTra}}) and relies on two $J^2$-operations instead of one.\\
\\
For all $l$, all operators $\sigma_{l,k_l}$ commute with each other and $\sigma_{l,k_l}^2=1$ holds for each of them. To simplify the notation during the proof, we abbreviate the $\sigma_{l,k_l}\equiv \sigma_l$. As a first step in the proof we can rewrite the exponential in $G_N$ as a product and use $\exp(-i \textstyle{\frac{\alpha}{2}}
\sigma_k \sigma_l)=\cos(\alpha/2)-i \sin(\alpha/2) \sigma_k \sigma_l$ to arrive at
\begin{equation}\label{Eq:ExpExp}
  \exp(-i \alpha D_N^2)=\exp(-i \textstyle{\frac{\alpha}{4}}
  N)\prod_{l=2}^{N} \prod_{k=1}^{l-1} \left( \cos(\alpha/2)- i
  \sin(\alpha/2) \sigma_k \sigma_l \right).
\end{equation}
For  $\alpha=\textstyle{\frac{\pi}{2}}$ we obtain
\begin{equation}\label{SpeExp}
  \exp(-i \textstyle{\frac{\pi}{2}} D_N^2)=\exp(-i \textstyle{\frac{\pi}{8}}
  N)\prod_{l=2}^{N} \prod_{k=1}^{l-1} \frac{1}{\sqrt{2}}\left(1 - i
 \sigma_k \sigma_l
 \right).
\end{equation}
All operations in the product on the right-hand side commute and can be decomposed into $\pi/2$-single qubit rotations and CNOT operations, which generate the Clifford group. Their action can therefore in principle be determined by using the techniques described in \cite{gottesman98}. Here we use a more elementary direct proof to get from Eq. (\ref{Eq:ExpExp}) to Eq. (\ref{Eq:UniTra}). We will use induction and treat $N$ even or odd separately.

\subsection{Even $N$}

Eq. (\ref{Eq:UniTra}) and Eq. (\ref{SpeExp}) are obviously equivalent for $N=2$. To show equivalence of equations (\ref{Eq:UniTra}) and
(\ref{SpeExp}) for $N'= N+2$ we re-express $U_{N+2}$ using Eq. (\ref{SpeExp}) as a product of $U_N$ times additional factors, and
substitute the right hand side of Eq. (\ref{SpeExp}) for $U_N$.
\begin{eqnarray}\label{RewExp}
U_{N+2}&=&\exp(-i \textstyle{\frac{\pi}{8}}
  (N+2))\prod_{l=2}^{N+2} \prod_{k=1}^{l-1} \frac{1}{\sqrt{2}}\left(1 - i
 \sigma_k \sigma_l
 \right) \nonumber\\
 &=&\frac{\exp(-i \pi/2)}{2}(1+i^{N+1} \sigma_1...\sigma_N
 -i \sigma_{N+1} \sigma_{N+2}+i^{N} \sigma_1...\sigma_{N+2})\nonumber\\
 &\times&\prod_{k=1}^N \textstyle{\frac{1}{2}}(1- i \sigma_k \sigma_{N+1})(1- i \sigma_k
 \sigma_{N+2}).
\end{eqnarray}
We can also verify that
\begin{eqnarray}\label{OthHan}
\textstyle{\frac{1}{2}}(1+i^{N+1} \sigma_1...\sigma_N
 -i \sigma_{N+1} \sigma_{N+2}+i^{N} \sigma_1...\sigma_{N+2}) \nonumber\\
\times \textstyle{\frac{1}{2}}(1+i^N \sigma_1...\sigma_N- \sigma_{N+1}\sigma_{N+2}+i^N
\sigma_1... \sigma_{N+2})\nonumber\\
 =\frac{\exp(i \pi/4)}{\sqrt{2}}\left(1+i^{N+3}
 \sigma_1...\sigma_{N+2}\right).
\end{eqnarray}
Therefore Eq. (\ref{Eq:UniTra}) is true for $N+2$ ($N$ even) if we can show that
\begin{equation}\label{AbRel}
\prod_{k=1}^{N} \textstyle{\frac{1}{2}}\left(1 - i
 \sigma_k a \right) \left(1 - i \sigma_k b
 \right)=\textstyle{\frac{1}{2}} \left(1+i^N \sigma_1...\sigma_N-
  a b+ i^N \sigma_1... \sigma_{N} a b\right),
\end{equation}
when $a^2=b^2=1$. This can be shown by another inductive proof. Again for $N=2$, relation (\ref{AbRel}) is easily verified. For $N$ replaced by $N+2$ we get
\begin{eqnarray}\label{AbRel2}
\prod_{k=1}^{N+2} \textstyle{\frac{1}{2}}\left(1 - i
 \sigma_k a \right) \left(1 - i \sigma_k b
 \right)\nonumber\\
 =\left[ \prod_{k=1}^{N} \textstyle{\frac{1}{2}}\left(1 - i
 \sigma_k a \right) \left(1 - i \sigma_k b
 \right)\right] \textstyle{\frac{1}{4}}\left(1-i\sigma_{N+1}(a+b)-a b \right)
 \left(1-i\sigma_{N+2}(a+b)-a b \right)\nonumber\\
=\textstyle{\frac{1}{4}}\left(1+i^N \sigma_1 ...\sigma_N -a b +i^N
\sigma_1 ...\sigma_N a b \right)\left(1-\sigma_{N+1}\sigma_{N+2}-a
b-\sigma_{N+1}\sigma_{N+2} ab \right)\nonumber\\
 =\textstyle{\frac{1}{2}} \left(1-
  a b+ (1+ a b) i^{N+2} \sigma_1... \sigma_{N}\right),
\end{eqnarray}
which completes the argument for $N$ even.
\subsection{Odd $N$}
For $N$ odd and $N\geq3$ the relation for $U_N$ corresponding to Eq. (\ref{SpeExp}) is
\begin{eqnarray}\label{OddExp}
 U_N=\exp(-i \textstyle{\frac{\pi}{2}} D_N) \exp(-i \textstyle{\frac{\pi}{2}} D_N^2)\nonumber\\
 =\exp(-i \textstyle{\frac{\pi}{8}}
  N)\left\{\prod_{l=2}^{N} \prod_{k=1}^{l-1} \frac{1}{\sqrt{2}}\left(1 - i
 \sigma_k \sigma_l
 \right)\right\} \prod_{m=1}^{N} \frac{1}{\sqrt{2}}\left(1 - i
 \sigma_m \right).
\end{eqnarray}
For $N=3$ the equivalence of Eq. (\ref{Eq:UniTra}) and Eq. (\ref{OddExp}) can be shown by explicit calculation. For $N'=N+2$ the proof proceeds similarly to the even case. We can first show that
\begin{eqnarray}\label{RewOdd}
 \exp(-i \textstyle{\frac{\pi}{2}} D_{N+2}) \exp(-i \textstyle{\frac{\pi}{2}}
 D_{N+2}^2)\nonumber\\
 =\frac{e^{-i \textstyle{\frac{\pi}{8}}
  }}{\sqrt{2}}\left(1-i^N \sigma_1...\sigma_N \right)
 \frac{1}{2}\left(1 -
 \sigma_{N+1}-\sigma_{N+2}- \sigma_{N+1} \sigma_{N+2}
 \right) \times \nonumber\\
 \times \prod_{k=1}^{N} \frac{1}{2}(1 - i
 \sigma_k \sigma_{N+1})(1-i \sigma_k \sigma_{N+2}).
\end{eqnarray}
Analogously to the even case, we can then prove by induction that for $a^2=b^2=1$
\begin{equation}\label{AbOdd}
(1-a-b-a b)\prod_{k=1}^{N} \textstyle{\frac{1}{2}}\left(1 - i
 \sigma_k (a + b)- a b
 \right)= \left(1- a b + (1+ a b) i^N \sigma_1...\sigma_N\right).
\end{equation}
Since
\begin{eqnarray}\label{OddOth}
\frac{e^{-i \textstyle{\frac{\pi}{8}}
  }}{\sqrt{2}}\left(1-i^N \sigma_1...\sigma_N \right)
 \frac{1}{2}\left(1 + i^N \sigma_1...\sigma_{N+2}+ i^N \sigma_1...\sigma_{N}- \sigma_{N+1} \sigma_{N+2}
 \right)= \nonumber\\
 \frac{e^{-i \textstyle{\frac{\pi}{8}}}}{\sqrt{2}}\left(1 - i^{N+2}
 \sigma_1... \sigma_{N+2}\right),
\end{eqnarray}
the proof is complete.
\section{Efficient stabilizer-code syndrome measurement}
\label{Sec:AppErr}
Stabilizer codes are a powerful tool for quantum error-correction and extensively discussed in the literature. An introduction to this subject and many of the early original references can be found in \cite{nielsen10}. Here we only need a few basic facts, namely that errors are detected by determining the eigenvalues $p~\epsilon ~\{-1,1\}$ of Pauli operator products that are designed to detect the absence or presence of a certain error. The Pauli product $P_N$  is then called a stabilizer and can be applied to a state. For a state $|p\rangle$ inside the code space, the eigenvalue, also called the syndrome is $P_N |p\rangle =+1 |p\rangle$, while measuring $P_N |p\rangle =-1 |p\rangle$ indicates that the code state has been compromised by a specific error that the stabilizer code is constructed to detect. By taking advantage of interference, we can determine the syndrome of the stabilizer $P_N$ acting on a string of $N$ qubits based on the generalized spin-spin interactions mediated by the associated operators $D_N$. Implementations require a constant number of multi-qubit operations and possibly a number of single qubit rotations that is linear in $N$, depending on  details of the implementation and the stabilizers in the code (see section \ref{Sec:Imp}). The read-out requires only one ancilla and does not disturb the code state $|p\rangle$ which can therefore remain encoded for further steps.\\
\\
We  assume that $N$ qubits are in an eigenstate $|p \rangle$ of a certain stabilizer $P_N$, $P_{N}|p \rangle = p |p \rangle$, with $p~\epsilon ~\{-1,1\}$. For the algorithm we add one ancilla qubit that we prepare in $| \uparrow \rangle$. In the following we assume that $N+1$ and $\textstyle{\frac{N+1}{2}}$ are even. Implementation of cases with $N+1$ odd will require additional single-qubit rotations and if $\textstyle{\frac{N+1}{2}}$ is not even, this will produce a different sign between the identity operator and the parity operator in Eq. (\ref{Eq:UnpPar}), which can be taken into account by changing the axis of the final rotation on the ancilla in Eq. (\ref{Eq:FinPul}) below. For our particular choice of $N+1$ even, $U_{N+1}$ simplifies to
\begin{equation}\label{Eq:UnpPar}
  U_{N+1}=\frac{e^{-i \textstyle{\frac{\pi}{4}}}}{\sqrt2} \left( 1+i^{N+2} \sigma_1...\sigma_{N+1}
  \right)=\frac{e^{-i \textstyle{\frac{\pi}{4}}}}{\sqrt2} \left( 1+ i P_{N}\sigma_{N+1}\right).
\end{equation}
We assume the ancilla at position $N+1$ was prepared in state $|\uparrow \rangle$ and first apply a $\pi/2$-rotation $R^{(1)}_y$ to the ancilla only and then $U_{N+1}$ as it appears on the right-hand side of Eq. (\ref{Eq:UnpPar}) to all $N+1$ qubits, including the ancilla, for which we choose $\sigma_{N+1}=\sigma_z$. This results in
\begin{eqnarray}\label{Eq:RotUn}
  U_{N+1} \left[|p \rangle (R^{(1)}_y | \downarrow \rangle)\right] &=&\frac{\exp[- i \pi/4]}{\sqrt{2}} \left(1+ i P_N \sigma_z \right)|p \rangle (| \uparrow \rangle+| \downarrow\rangle)\nonumber\\
  &=&|p \rangle \frac{\exp[-i \pi/4]}{\sqrt{2}}\left( \frac{1+ i p}{\sqrt{2}} | \uparrow\rangle + \frac{1- i p}{\sqrt{2}}  |\downarrow\rangle \right).
\end{eqnarray}
A final rotation $(R^{(1)}_x)^\dag$ of the ancilla qubit around the $x$-direction will produce the state
\begin{equation}\label{Eq:FinPul}
  (R^{(1)}_x)^\dag U_{N+1} \left[|p \rangle (R^{(1)}_y | \downarrow \rangle)\right] = |p \rangle\left(\frac{1+p}{2}| \uparrow \rangle+ \frac{1-p}{2}| \downarrow\rangle \right).
\end{equation}
The syndrome of $|p \rangle$, is now encoded in the state of the ancilla which will be $|\uparrow\rangle$ for $p=1$ or $|\downarrow\rangle$ for $p=-1$ and can be read out
without disturbing the code state $|p \rangle$.\\
\\
If the input state to this pulse sequence is not an eigenstate of the stabilizer $P_N$, finding the ancilla in the state corresponding to $p$ will project the initial state into a superposition of eigenstates $|S_p\rangle$ with eigenvalue $p$. If $p=1$, $|S_1\rangle$ contains at least one code state of the stabilizer code. Otherwise, we can take note that $p=-1$, which means that  $|S_{-1}\rangle$ contains at least one state of the code rotated by one of the errors that yield $p=-1$ when $P_N$ is applied. After applying all error operators and learning their syndromes, the resulting state can be recovered into the code space by applying the correction operation compatible with all the syndrome results we have found. To exploit this for code-state preparation, we can initialize the $N$ qubits in an equal superposition of all states by applying a collective $\pi/2$-rotation $R^{(N)}_y$ and then successively measure all stabilizers $P_N$ that are error operators of the code. We keep track of all instances of measuring $p=-1$ and eventually apply the correction operation corresponding to the set of syndromes we have found. After this correction, the resulting state $|S^{(f)}_1\rangle$ must be a superposition of code states and can be projected into a certain encoded qubit state by, for example, finally measuring the eigenvalue of the {\it encoded} operator $\bar{\sigma}_z$ which is implemented by another stabilizer $P_{\bar{\sigma}_z}$. In this way, an encoded qubit state can be efficiently produced in $N$ steps. In principle this scheme can be executed several times to increase the probability of obtaining a code state, even if not all operations and measurements are perfect.\\
\\  
In the special case of measuring the parity operator $P_N$, by using the uniform interaction $D_N=J_z$, the parity can be determined with three ($N$ even) or four operations ($N$ odd) and revealed by one measurement on the ancilla qubit, independent of the size of $N$. This constant depth parity-finding algorithm is in contrast with parity determination algorithms on conventional computers which require polynomial sized circuit families of close to $\log{N}$ depth for computing parity \cite{heydon90}. The quantum version therefore provides a genuine advantage even for relatively small $N$ and falls into a family of constant depth parity circuits discussed in \cite{hoyer05}. 
\section{Implementations in ion trap systems}\label{Sec:Imp}
In many codes, for example Calderbank-Shore-Steane (CSS) codes \cite{nielsen10}, stabilizers are products of only one Pauli-operator, either $\sigma_{l,x}$ or $\sigma_{l,y}$, and  the identity $\sigma_{l,i}$. For such codes, stabilizer measurements with just M{\o}lmer and S{\o}rensen type interactions with either $\phi=0$ for $\sigma_{l,\phi}=\sigma_{l,x}$ or $\phi=-\pi/2$ for $\sigma_\phi=\sigma_{l,y}$ can be used. The third Pauli operator is not required because a $\sigma_{l,z}$ (phase-flip) error can be thought of as sequentially occurring $\sigma_{l,x}$ and $\sigma_{l,y}$ (bit-flip) errors.\\
\\
The conceptually simplest way to realize identities could be to momnetarily physically remove all qubits that are acted on by $\sigma_{l,i}$ within $D_N$ (so no physical operation on that qubit is required), for example by transporting them to a location away from where interactions are applied in a multi-zone architecture \cite{wineland98,kielpinski02}. This effectively reduces the original stabilizer to $P_{N'}$ with $N' < N$ and all elements in the stabilizer $P_{N'}$ are Pauli-matrices. We can then either apply M{\o}lmer and S{\o}rensen type interactions with two different phases, or, for non-CSS codes that may require more complicated Pauli products, rotate each remaining qubit individually by $r_l$ such that $\sigma_{l,k_l} =r^\dag_l \sigma_{l,z} r_l$. If we denote the operator that applies such individual rotations to all remaining qubits as $R_{k,N'}=\prod_{l=1}^{N'} r_l$ we have $D_{N'}=R^\dag_{k,N'} J_z R_{k,N'}$ and can rewrite any integer power $m$ of $D_{N'}$ as 
\begin{equation}
(D_{N'})^m= R^\dag_{k,N'} J_z (R_{k,N'} R^\dag_{k,N'}) J_z (R_{k,N'} R^\dag_{k,N'})... J_z R_{k,N'} =R^\dag_{k,N'} (J_z)^m R_{k,N'}.
\end{equation}
Because this reasoning can be applied to every term in the sum defining the operator-exponential function, we can initially apply $R_{k,N'}$, then $U_z$ and then undo the initial rotation with $R^\dag_{k,N'}$, which implies $G_{N'}=R^\dag_{k,N'} U_z R_{k,N'}$. After this we can recombine all $N$ qubits and have effectively implemented $G_N$ up to a global phase, because $G_N=\exp[i \alpha (N-N')]G_{N'}$. Any other uniform $J^2$ interaction $U_\phi$ can also be used for the uniform coupling of all $N'$ qubits, after modifying the initial and final rotations to match the direction specified by $\phi$.\\
\\
Separation of qubits is not required in an architecture as described in \cite{nebendahl09}, where, in a minimal construction, M{\o}lmer and S{\o}rensen $J_x^2$ or $J_y^2$ type interactions that act globally on all $N$ qubits can be supplemented by individually addressed $\sigma_{l,z}$ rotations, applied to only the qubits that are supposed to not partake in the $J^2$-interaction. The individual rotations are realized by AC-Stark shifts and refocus each of the addressed qubits in such a way that their state is unchanged by the total operation. Depending on the particular structure of the error-correction code, a more optimal sequence in the sense defined in \cite{nebendahl09} may also exist, but will be hard to find using the numerical optimization methods described in this work as the state space of the code increases in size. In any case, this type of implementation should be particularly efficient if all stabilizers contain the same Pauli-matrix apart from the individually refocused group of qubits, as in CSS codes.\\
\\
Alternatively, we can use an array of tightly focused laser beams that address ions individually \cite{linke17} and allow for precise definition of individual operations $\sigma_{l,k_l}$ on each ion (with $\sigma_{l,i}$ corresponding to not turning the individual beam on for the $l$-th ion). With such a setup, any $D^2_N$ interaction can be implemented directly and executed in a constant number of parallel operations (constant depth).
\section*{Acknowledgements}
This work was mostly developed in the environment of the NIST Ion Storage Group, which therefore owns a great deal of the credit. We would like to thank Jim Bergquist, John Bollinger, James Chou, David Hume, Wayne Itano, David Leibrandt and  Andrew Wilson as well as all the post-docs, grad and summer students and the administrative and technical staff of the Time and Frequency Division at NIST. In addition, we gratefully acknowledge numerous discussions and useful advice from Manny Knill and the members of his group.\\
\\
This work was supported by funding from the Office of the Director of National Intelligence (ODNI), the Intelligence Advanced Research Projects Activity (IARPA) and the NIST Quantum Information Program.\\
\\
This paper is a contribution of NIST and is not subject to US copyright.%
%

%
\end{document}